\documentclass[10pt,letterpaper]{article}
\usepackage[top=0.85in,left=2.75in,footskip=0.75in]{geometry}

\usepackage{changepage}

\usepackage[utf8]{inputenc}

\usepackage{textcomp,marvosym}

\usepackage{fixltx2e}

\usepackage{amsmath,amssymb}

\usepackage{cite}

\usepackage{nameref,hyperref}

\usepackage[right]{lineno}

\usepackage{microtype}
\DisableLigatures[f]{encoding = *, family = * }

\usepackage{rotating}

\usepackage{subfigure}

\usepackage{xcolor}

\usepackage{parskip}

\raggedright
\usepackage[aboveskip=1pt,labelfont=bf,labelsep=period,justification=raggedright,singlelinecheck=off]{caption}

\makeatletter
\renewcommand{\@biblabel}[1]{\quad#1.}
\makeatother

\date{}

\usepackage{lastpage,fancyhdr,graphicx}
\usepackage{epstopdf}
\fancyheadoffset[L]{2.25in}
\fancyfootoffset[L]{2.25in}
\begin{document}
\vspace*{0.35in}

\begin{flushleft}
{\Large
\textbf\newline{\bf Residual sweeping effect in turbulent  particle pair diffusion in a Lagrangian diffusion model}
}
\newline
\\
Malik, Nadeem A. 
\\
\bigskip
Department of Mathematics and Statistics,\\
King Fahd University of Petroleum and Minerals,\\
P.O. Box 5046, Dhahran 31261, Saudi Arabia
\\
\bigskip

E-mail: namalik@kfupm.edu.sa, nadeem\_malik@cantab.net

\end{flushleft}

\setlength{\parskip}{5pt}
\setlength\parindent{0pt}

\section*{Abstract}
Thomson, D. J. \& Devenish, B. J. [{\em J.~Fluid Mech.} 526, 277 (2005)] and others  have suggested that sweeping effects make Lagrangian properties in Kinematic Simulations (KS), Fung et al [Fung J. C. H., Hunt J. C. R.,  Malik N. A. \& Perkins R. J. {\em J.~Fluid Mech.} 236, 281 (1992)], unreliable. Here it is shown through a novel analysis based upon analysing pairs of particle trajectories in a frame of reference moving with the large energy containing scales of motion that the  normalized integrated error $e^I_K$ in the turbulent pair diffusivity ($K$) due to the sweeping effect  decreases with increasing pair separation ($\sigma_l$), such that $e^I_K\to 0$ as $\sigma_l/\eta\to \infty$; and $e^I_K\to \infty$ as $\sigma_l/\eta\to 0$. $\eta$ is the Kolmogorov turbulence microscale. There is an intermediate range of separations $1<\sigma_l/\eta< \infty$ in which the error $e^I_K$ remains negligible. Simulations using KS shows that in the swept frame of reference, this intermediate range is large covering almost the entire inertial subrange simulated, $1<\sigma_l/\eta< 10^5$, implying  that the deviation from locality observed in KS therefore cannot  be atributed to sweeping errors and could be real. This is important for pair diffusion theory and modeling.

\bigskip
\noindent PACS numbers:  47.27.E?, 47.27.Gs, 47.27.jv, 47.27.Ak, 47.27.tb, 47.27.eb, 47.11.-j 

\noindent Keywords: Turbulence, diffusion, particle pair, pair diffusivity, Kinematics Simulations, numerical analysis, Lagrangian, sweeping effect



\section*{Introduction}

Turbulent particle pair diffusion has attained somewhat of an iconic status in the turbulence community, many researchers having addressed this topic over the decades. Nevertheless, most if not all theories of turbulent particle pair diffusion in homogeneous turbulence with extended inertial ranges have been based upon the hypothesis of locality since Richardson in 1926 \cite{Richardson1926}, and Obukov in 1941  \cite{Obukhov1941}.

Richardson pioneered this field and introduced the idea of a scale dependent pair diffusivity as the fundamental quantity of interest in turbulent pair diffusion studies. The turbulent pair diffusivity is defined as,  
\begin{eqnarray}
   K(l) = {1\over 2}{d\langle l^2\rangle\over dt} = \langle {\bf l}\cdot {\bf v}(l)\rangle, 
\end{eqnarray}
where ${\bf l}(t)$ is the pair displacement vector at time $t$, $l=|{\bf l}|$, ${\bf v}(l)$ is the pair relative velocity, and $\langle\cdot\rangle$ is the ensemble average over all particle pairs.
  
The locality hypothesis can readily be applied to generalized power law spectra of the type, $E(k)\sim k^{-p}$, for $1<p\le 3$; the pair diffusivity then scales like $K(l,p)\sim \sigma_l^{\gamma^l_p}$ with $\gamma^l_p =(1+p)/2$, \cite{Morel1974}, where $\sigma^2_l=\langle l^2\rangle$.  For Kolmogorov turbulence $p=5/3$, this gives the well known Richardson scaling $K\sim \sigma_l^{4/3}$, which is equivalent to $\langle l^2\rangle\sim t^3$ \cite{Obukhov1941}. 

Kinematic Simulations (KS) \cite{Kraichnan1970,Fung1992} has often been used to investigate turbulent pair diffusion, even though it does not yield the assumed locality scaling; for $p=5/3$, KS gives $\gamma_{Kol}\approx 1.53 >4/3$. For this reason, it has been widely assumed that KS must be in error. Thomson \& Devenish \cite{TD2005} argue that the turbulent pair diffusivity must scale like,  
\begin{eqnarray}
  K(l(t))\sim S(l) \tau_s(l(t)),
\end{eqnarray}

where $S(l)$ is the structure function {of the turbulence velocity field} and $\tau_s(l)$ is an effective time scale of velocity increments. In real turbulence, 
assuming locality scaling for $S(l)\sim l^{2/3}$ and for $\tau_s(l)\sim l^{2/3}$ leads to Richardson's classical scaling $K(l)\sim \sigma_l^{4/3}$, where we evaluate $K$ at typical values of $l$, namely $\sigma_l=\langle l^2\rangle^{1/2}$ which is commonly assumed in these studies.

Thomson \& Devenish argue that in KS because of the lack of true dynamical sweeping, the time scale must be a sweeping time scale, namely $\tau_s\sim l/U_s$  which is a time scale for the large scales flow to cut through smaller local eddies, $U_s$ being the sweeping velocity scale. This leads to, $K\sim \sigma_l^{5/3}$. Even when the rms turbulence velocity $u'$ is taken instead of $U_s$, they obtained $K\sim \sigma_l^{14/9}$. 

They concluded that whereas locality is true in real turbulence, it is not true in KS.  In turbulence the large energy containing eddies carry the smaller eddies, but in KS as there is an absence of true dynamics the large scales force the fluid particles to cut through the smaller eddies in an unphysical manner, a view supported by Nicolleau \& Nowakowski \cite{Nicolleau2011}, and Eyink \& Benveniste \cite{Eyink2013}.

However, Thomson \& Devenish's scaling argument leading to equation (2) adresses only the scaling laws for $K$, but does not quantified the actual errors in the diffusivity $K$ in KS -- is it large or small ? It is prudent, therefore, to seek an alternative, a more analytic, approach to address this question, which is the main concern of this work.

Here we re-examine the sweeping effect in KS with a view of quantitfying the error in the KS pair diffusivity $K^s$ compared to the physical pair diffusivity $K$. For  this purpose, we focus upon the {\em differences} in the relative velocities along pairs of particle paths in  the {\em sweeping frame of reference}. This frame of reference accounts for the physical sweeping effect of the largest energy containing scales; but a residual sweeping effect still remains due to the largest inertial range eddies sweeping the smaller inertial range eddies.

Consider Fig. 1 which shows a particle pair with separation ${\bf l}$ in the inertial subrange  being swept by a large scale flow. {A real fluid particle pair will be swept by the physical velocity field ${\bf u}$} and will follow certain particle paths; but a KS flow will transport the pair along neighbouring particle paths due to an additional KS sweeping motion,  {${\bf u}^s$}, and thereby force the particles to cut through local flow structures. 

The large scale physical sweeping velocity are assumed not to affect the relative motion of particles in a pair in the inertial subrange. The critical question is, are the deviations from the physical trajectories induced by KS in the pair diffusion process large or small?

In the ensuing analysis, it is the error between the KS and the physical pair diffusivities, $|K^s-K|$, that will be calculated. We will consider generalised power law energy spectra, $E(k)\sim k^{-p}$, because the analysis must be valid for all such power spectra and this will add weight to the results and conclusions that can be drawn from this work if validated over the wholw range of $p$ considered.

The main questions of interest are, is the KS sweeping error large or small, and in what range of separations? These questions are addressed first through a novel mathematical analysis focussing upon pairs of neighbouring particle trajectories. This is then verified against simulations using KS with very large inertial subranges. 

{In Section 1, we derive an expression for the error in the pair diffusivity in KS flows by analysing neighbouring trajectories in the swept frame of reference. In Section 2, the KS method is discussed and simulation results presented. In the final Section 3, we discuss the results and its implications for theory and modeling.}

\section{The normalized error in the pair diffusivity}

\subsection{The numerical timestep error} 

In the swept frame of reference, the relative motion of a particle pair in the inertial subrange is unaffected by the sweeping action itself. This can be mimicked in KS by setting $E(k)=0$ for $k<k_1$. However,  there still remains a residual sweeping caused by the largest of the inertial scales sweeping the scales local to the pair separation. 

Consider an ensemble of particle pairs released in a field of homogeneous turbulence at time $t=0$ with some small initial separation $l_0$. At some time $t$ later, the ensemble average of the separation is assumed to be well  inside the inertial subrange and the relative motions are independent of $l_0$ \cite{Batchelor1952}.

Consider the particles in one of these pairs,  labeled $1$ and $2$, as shown in Fig. 1. The particle locations are ${\bf x}_1(t)$ and ${\bf x}_2(t)$ respectively at time $t$; and the pair displacement is ${\bf l}(t)={\bf x}_2-{\bf x}_1$, and $l(t)=|{\bf x}_2-{\bf x}_1|$.  The physical flow is ${\bf u}({\bf x},t)$. All quantities are assumed at time $t$ unless otherwise stated. 

At time $t$ the {\em additional} (or {\em residual}) KS sweeping flow ${\bf u}^s({\bf x},t)$ is 'swiched on' -- this is not to be confused with the total KS velocity field which is $({\bf u}+{\bf u}^s)({\bf x},t)$, see Fig. 1.

The physical flow ${\bf u}({\bf x},t)$ transports the particles to ${\bf x}_1(t^*)$ and ${\bf x}_2(t^*)$ respectively at the next time step $t^*=t+dt$; while the KS flow $({\bf u}+{\bf u}^s)({\bf x},t)$  transports the particles to ${\bf x}^s_1(t^*)$ and ${\bf x}^s_2(t^*)$ respectively. Note that ${\bf l}^s={\bf x}^s_2-{\bf x}^s_1$, and $l^s = |{\bf x}^s_2-{\bf x}^s_1|$.  

The superscript ${}^*$ will refer to quantities at time $t^*$, e.g. $l^*=l(t+dt)$. The superscript ${}^s$ will refer to quantities related to the KS residual sweeping, e.g. $l^s(t^*)=l^s(t+dt)$.

The following quantities are defined:

${\bf u}={\bf u}({\bf x},t)$  \hbox{is the physical fluid velocity field}\\
${\bf u}^s={\bf u}^s({\bf x},t)$ \hbox{is the  additional (residual) sweeping velocity field}\\
${\bf v}({\bf l})={\bf u}({\bf x}_2)-{\bf u}({\bf x}_1)$ \hbox{\rm is the physical relative velocity}\\
${\bf v}^s({\bf l})={\bf u}^s({\bf x}_2)-{\bf u}^s({\bf x}_1)$ \hbox{is the additional (residual) relative velocity}\\
${\bf \tilde u}=({\bf u}+{\bf u}^s)({\bf x},t) $ \hbox{is the total KS velocity}\\
${\bf \tilde v}({\bf l}^s)={\bf v}({\bf l}^s)+{\bf v}^s({\bf l}^s) $ \hbox{is the total KS relative velocity}

Simplifying the notation as much as possible, e.g. ${\bf u}_2={\bf u}({\bf x}_2,t)$, and ${\bf u}^*_2={\bf u}({\bf x}_2,t+dt)$, yields

\begin{eqnarray}
{\bf x}^s_1(t^*)&=&{\bf x}_1(t^*)+ {\bf u}^s_1(t) dt\\
{\bf x}^s_2(t^*)&=&{\bf x}_2(t^*)+ {\bf u}^s_2(t) dt\\
{\bf l}^{s*}&=&{\bf x}^s_2(t^*)-{\bf x}^s_1(t^*) \nonumber\\
 &=&{\bf l}(t^*)+ ({\bf u}^s_2- {\bf u}^s_1)dt \nonumber\\
 &=&{\bf l}^* + {\bf v}^s({\bf l})dt 
\end{eqnarray}

${\bf \tilde v}(l^{s*})$ is calculated at the new KS swept particle locations. Using Taylor expansions wherever necessary, assuming that the velocity fields are at least twice differentiable in space and at least once in time,

\begin{eqnarray}
{\bf \tilde v}(l^{s*})&=&({\bf u + u^s})({\bf x}^s_2(t^*))-({\bf u+u^s})({\bf x}^s_1(t^*)) \nonumber\\
&=& {\bf u}({\bf x}^s_2(t^*))-{\bf u}({\bf x}^s_1(t^*)) 
+ {\bf u}^s({\bf x}^s_2(t^*))-{\bf u}^s({\bf x}^s_1(t^*)) \nonumber\\
&=& {\bf v}({\bf l}^{*}) + {\bf v}^s({\bf l}^{*}) + \nonumber\\
&&\left({{\bf u}^s_2\cdot\nabla}{\bf u}_2(t^*)-{{\bf u}^s_1\cdot\nabla}{\bf u}_1(t^*)\right)dt +\nonumber\\
&&\left({{\bf u}^s_2\cdot\nabla}{\bf u}^s_2(t^*)-{{\bf u}^s_1\cdot\nabla}{\bf u}^s_1(t^*)\right)dt \nonumber\\ &&+O(dt^2)
\end{eqnarray}

\begin{figure*}
\begin{center}
\mbox{\subfigure{\includegraphics[width=14cm]{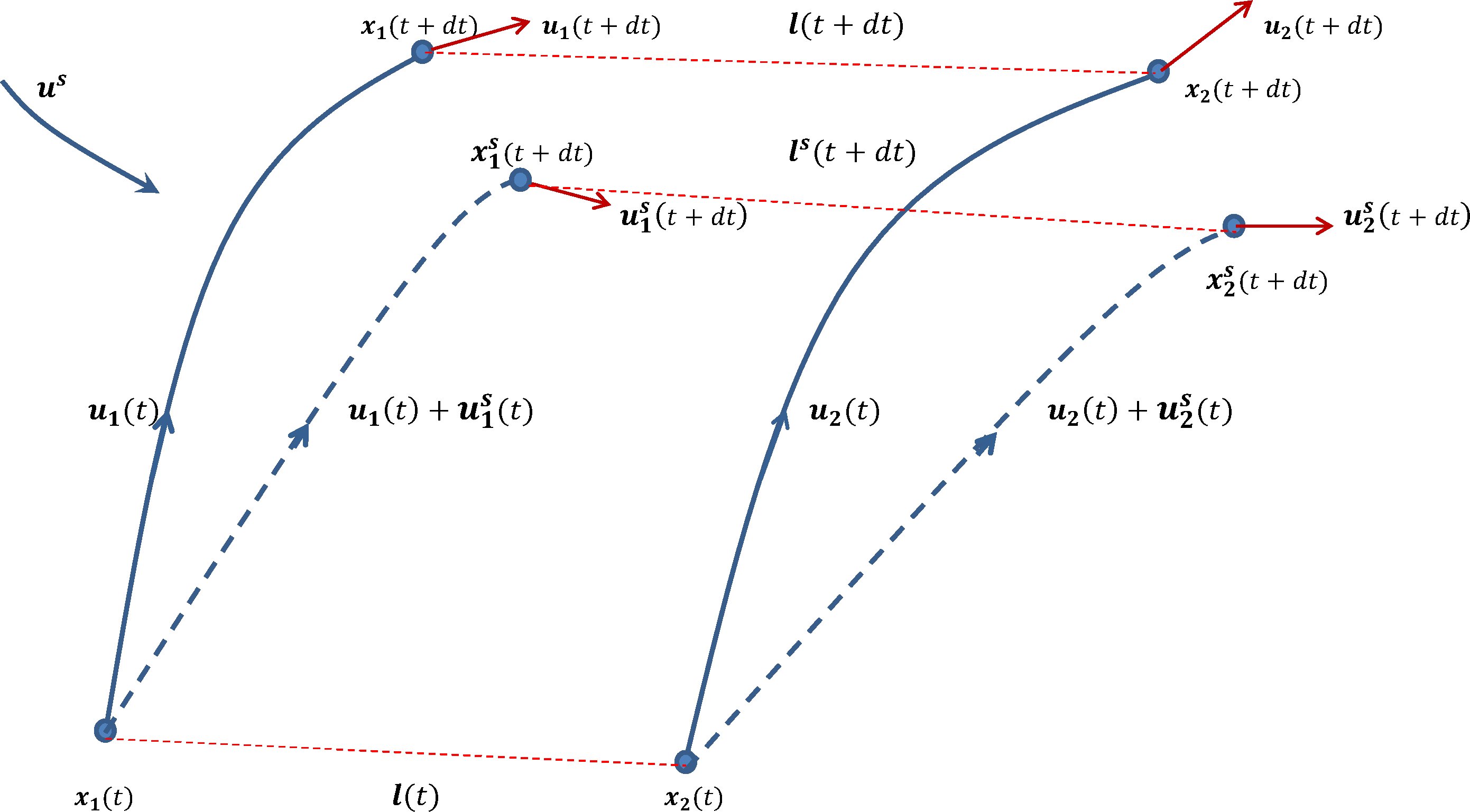}}}
\caption{\label{fig1} {Schematic diagram illustrating the system discussed in the text. The locations of two nearby particles, labelled 1 and 2, are located at ${\bf x}_1(t)$ and ${\bf x}_2(t)$ respectively,  with turbulence velocities ${\bf u}_1(t)$ and ${\bf u}_2(t)$, at time $t$; their separation is $l(t)=|{\bf x}_2-{\bf x}_1|$. They are transported with velocities ${(\bf u_1 + u_1^s)}(t)$ and ${(\bf u_2 + u_2^s)}(t)$ respectively to the new locations ${\bf x}^s_1(t)$ and ${\bf x}^s_2(t)$ at the next time step $t+dt$, as shown.}}
\end{center}
\end{figure*}

The pair diffusivity at time $t^*$ is, $K^*=\langle{\bf l}^*\cdot {\bf v}(l^*) \rangle$ {-- we ignore constants of proportionality, like $2$, because we are interested only in the power scalings in this work.} The KS equivalent is  $K^{s*}=\langle{\bf l}^{s*}\cdot {\bf \tilde v}(l^{s*}) \rangle$. Using equations (5) and (6) and ignoring terms of order $dt^2$ and higher,

\begin{eqnarray}
K^{s*}&\approx&
\langle{\bf l}^*\cdot{\bf v}({\bf l}^*)\rangle + 
\langle{\bf l}^*\cdot{\bf v}^s({\bf l}^*)\rangle + \nonumber\\
&&\langle{\bf l}^*\cdot\left({{\bf u}^s_2(t)\cdot\nabla}{\bf u}_2(t^*)-{{\bf u}^s_1(t)\cdot\nabla}{\bf u}_1(t^*)\right)\rangle dt +\nonumber\\
&&\langle{\bf l}^*\cdot\left({{\bf u}^s_2(t)\cdot\nabla}{\bf u}^s_2(t^*)-{{\bf u}^s_1(t)\cdot\nabla}{\bf u}^s_1(t^*)\right)\rangle dt + \nonumber\\
&& \langle {\bf v}^s({\bf l})\cdot{\bf v}({\bf l}^*)\rangle dt + 
     \langle {\bf v}^s({\bf l})\cdot{\bf v}^s({\bf l}^*)\rangle dt
\end{eqnarray}

The timestep error between the KS and physical diffusivities for a given timestep $dt$ is,   $E_K=|K^{s*}-K^*|$. Using the expansion ${\bf u}^s({\bf x}_2(t))\approx {\bf u}^s_1+
{\bf l}\cdot \nabla {\bf u}^s_1$ in equation (7), yields

\begin{eqnarray}
E_K &\approx&  
\langle{\bf l}^*\cdot{\bf v}^s({\bf l}^*)\rangle + \nonumber\\
&& \langle {\bf l}^* \cdot ({\bf u}^s_1\cdot \nabla) {\bf v}({\bf l}^*)\rangle dt + \nonumber\\
&& \langle {\bf l}^* \cdot ({\bf u}^s_1\cdot \nabla) {\bf v}^s({\bf l}^*)\rangle dt + \nonumber \\
&&\langle {\bf l}^* \cdot ({\bf l}\cdot \nabla){\bf u}^s_1\cdot \nabla {\bf u}_2(t^*)\rangle dt + \nonumber\\
&&\langle {\bf l}^* \cdot ({\bf l}\cdot \nabla){\bf u}^s_1\cdot \nabla {\bf u}^s_2(t^*)\rangle dt +\nonumber\\
&& \langle {\bf v}^s({\bf l})\cdot{\bf v}({\bf l}^*)\rangle dt + \nonumber\\
&&   \langle {\bf v}^s({\bf l})\cdot{\bf v}^s({\bf l}^*)\rangle dt
\end{eqnarray}

The time scale of the sweeping $T_s$ is much larger than the local time scale of the pair separation. Hence, in the last four terms ${\bf l}(t)$ is replaced by ${\bf l}(t^*)$ without affecting their magnitudes or scalings (the associated errors are $\sim O(dt^2)$ which is neglected).  

All the terms in equation (8) are now evaluated at the same time $t^*$, so without loss of generality $t^*$ is replaced by $t$ and the superscritp '$^*$' is dropped. The subscript '$_1$' is also dropped because of homogeneity. Equation (8) now simplifies to,

\begin{eqnarray}
E_K &\approx&  
\langle{\bf l}\cdot{\bf v}^s\rangle + \nonumber\\
&&\langle {\bf l} \cdot ({\bf u}^s\cdot \nabla) {\bf v}\rangle dt + \nonumber\\
&&\langle {\bf l} \cdot ({\bf u}^s\cdot \nabla) {\bf v}^s\rangle dt +\nonumber\\
&&\langle {\bf l} \cdot
({\bf l}\cdot \nabla){\bf u}^s\cdot \nabla {\bf u}_2\rangle dt + \nonumber\\
&&\langle {\bf l} \cdot
({\bf l}\cdot \nabla){\bf u}^s\cdot \nabla {\bf u}^s_2\rangle dt +\nonumber\\
&& \langle {\bf v}^s\cdot{\bf v}\rangle dt +  \nonumber\\
&&\langle ({\bf v}^s)^2\rangle dt 
\end{eqnarray}

It is reasonable to assume that the (residual) sweeping flow field ${\bf u}^s$, which is caused by the largest of the inertial range eddies, is close to uniform across small distances, and therefore the relative velocities across local scales $l$ that it induces is small. 
However, gradients of the relative velocity ${\bf v}(l)$ itself can be large. The magnitude $u^s=|{\bf u}^s|$ is assumed large compared to $v(l)=|{\bf v}(l)|$, and also as compared to $v^s(l)=|{\bf v}^s(l)|$. $u^s$ scales differently to $v(l)$.

$v(l)$ also scales differently to $v^s(l)$, the former being governed by inertial range turbulence scaling, and the latter by {\em differences} in the residual sweeping velocity across a small distance $l$. This can be seen clearly in the limit of uniform (parallel) sweeping flow, where the ${\bf v}(l)$ is unaffected, but ${\bf v}^s(l)={\bf 0}$ and all the terms on the right hand side in equation (9) are  zero except for the second term. This indicates that the second term  in equation (9) makes the dominant contribution to the error.

Consider generalized energy spectra of the form $E(k)=\varepsilon^{2/3}L^{5/3-p}k^{-p}$, for  $k_1\le k\le k_\eta$ and for $1<p\le 3$, and with $k_\eta/k_1 \gg 1$. In the swept frame of reference, $E(k)=0$ for $k<k_1$. The rate of energy dissipation is $\varepsilon\sim U^3/L$, where $U$ is the velocity scale in the energy containing scales.  

The previous discussion implies that  $|{\bf \nabla v}^s| \ll \left({U\over L}\right)$  and therefore, 
\begin{eqnarray} v^s (l) \ll \left({l\over L}\right)U. \end{eqnarray}
The energy in turbulent inertial scales local to $l$ is, $v^2(l)\sim E(1/l)/l$, and therefore,
\begin{eqnarray}
  v(l) &\sim& \left({l\over L}\right)^{{p-1\over 2}} U\\
  {\rm and,}\ \ |{\bf\nabla} v({l})| &\sim& \left({l\over L}\right)^{{p-3\over 2}} 
                      {U\over L}
\end{eqnarray}

It is usual to assume the scaling  $l\sim \sigma_l$ as previously mentioned. Then, the second term in equation (9) is given by,

\begin{eqnarray}
  E_2 &=& \langle {\bf l} \cdot ({\bf u}^s\cdot \nabla) {\bf v}\rangle dt 
\sim\left({\sigma_l\over L}\right)^{{p-1\over 2}} Uu^s(l) dt,
\end{eqnarray}

All the other terms in equation (9), labeled respectively $E_1, E_3, ... , E_7$,  scale proportional to $u^s$ or are much smaller, and this leads to the following estimates,

\begin{eqnarray}
 {E_1\over E_2 } &\ll&  \left({\sigma_l\over L}\right)^{5-p\over 2}{L\over Udt }, \quad \nonumber\\
 {E_3\over E_2 } &\ll&  \left({\sigma_l\over L}\right)^{3-p\over 2}  \quad \nonumber\\
 {E_4\over E_2 } &\ll&  \left({\sigma_l\over L}\right)^{5-p\over 2} \quad \nonumber\\
 {E_5\over E_2 } &\ll&  \left({\sigma_l\over L}\right)^{5-p\over 2}  \quad \nonumber\\
 {E_6\over E_2 } &\ll&  \left({\sigma_l\over L}\right)^{1} \quad \nonumber\\
 {E_7\over E_2 } &\ll&  \left({\sigma_l\over L}\right)^{5-p\over 2}.
\end{eqnarray}

All of these ratios are small for $\sigma_l/L < 1$ and $1<p\le 3$.  This is true even in the expression for $E_1/E_2$ because the factor $L/Udt$ is subdued by the very small factor in the brackets. It is reasonable to conclude that the 2nd term in equation (9) is dominant and therefore $E_K\approx E_2$.  

To estimate $E_2$ itself, an estimate for $u^s(l)$ is needed. The inertial subrange contains only about $1\%$ of the the total energy in the turbulence across the entire wavenumber range, $0<k<\infty$, such as in a von Karman spectrum $E_{vk}$,  i.e. $E_{ks}\approx E_{vk}/100$. This means that the root mean square turbulent velocity in the inertial range is approximately, $u_{ks}\approx U/10$, where $E_{vk}\sim U^2$. 

The wavenumbers that contribute to the sweeping of particle pairs at separation $\sigma_l$ are in the range $k_1\le k< k_l$, where $k_l\sim 1/\sigma_l$. In the KS sweeping frame of reference, the larger inertial scales are the sweeping scales, and the energy in these scales is approximately,

\begin{figure}
\begin{center}
\mbox{\subfigure{\includegraphics[width=12cm]{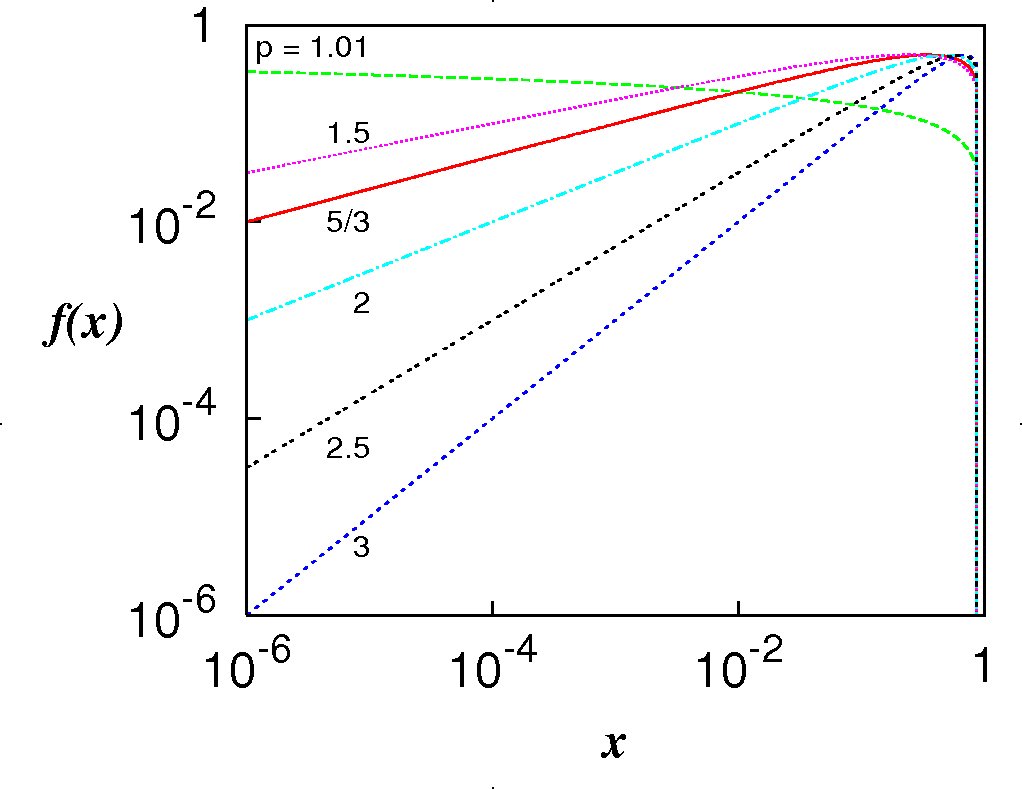}}}
\caption{
\label{fig2} The function $f(x)=\displaystyle{\sqrt{1-x^{p-1}} x^{(p-1)/2}}$,  $0< x\le 1$, for selected powers $1<p\le 3$.}
\end{center}
\end{figure}

\begin{eqnarray}
  (u^s)^2 &\sim& \int_{k_l}^{k_1} E(k)dk 
                  = \int_{k_l}^{k_1} \alpha_k \varepsilon^{2/3} L^{5/3-p}k^{-p} dk
\end{eqnarray}

and using $\varepsilon \sim (u_{ks})^3/L \sim (U/10)^3/L$, this gives

\begin{eqnarray}
   u^s &\approx& {U\over 10} \sqrt{1-\left( {\sigma_l\over L}\right)^{p-1}}
\end{eqnarray}

Using this in equation (), the residual error between the physical and the KS pair diffusivities in the inertial subrange of pair separations in the swept frame of reference, per unit timestep (retaining $E_K$ to represent this quantity), is 

\begin{eqnarray}
  E_K(p)\approx E_2 &=& C_k  \sqrt{1-\left( {\sigma_l\over L}\right)^{p-1}}
\left({\sigma_l\over L}\right)^{{p-1\over 2}} {U^2\over 10}.
\end{eqnarray}

$C_k$ is the constant of proportionality, which can depend upon $p$. 

For $p=5/3$, the residual error per unit timestep is,

\begin{eqnarray}
  E_K(5/3)&\approx& C_k  \sqrt{1-\left( {\sigma_l\over L}\right)^{2/3}}
\left({\sigma_l\over L}\right)^{1/3} {U^2\over 10}.
\end{eqnarray}

As $p\to 1$, $E_K\to E_K(1) \approx C_kU^2/L\approx Constant$ for $\sigma_l/L<1$, but is nearly zero close to $\sigma_l/L = 1$. In this limit, $p\to 1$, the pair diffusion is strongly local and is not affected by long  range sweeping. 

For $p=3$, the residual error per unit timestep is negligibly small for $\sigma_l/L<1$,

\begin{eqnarray}
  E_K(3)&\approx& C_k  \sqrt{1-\left( {\sigma_l\over L}\right)^{2}}
\left({\sigma_l\over L}\right) {U^2\over 10} \ll E_k(5/3).
\end{eqnarray}

In this limit, nearly all the energy is contained in the largest scales and inertial subrange scaling is no longer applicable.

\subsection{The integrated error} 

Equations (17) provides a way of estimating an upper bound for the integrated residual error. Fig. 2 shows the log-log plots of the factor $f(x) =\sqrt{1-x^{p-1}}x^{(p-1)/2}$  for selected powers in the range $0<p\le 3$. The range over which $\sqrt{1-x^{p-1}}\approx 0$ is very short and close to $x=1$; but over the rest of the range $\sqrt{1-x^{p-1}}\approx 1$. Hence, to a good approximation, $f(x)<x^{(p-1)/2}$ for all $x\le 1$, and using this in equation (17) with $x=\sigma_l/L$ yields,

\begin{eqnarray}
  E_K(p)&<&  C_k{U^2\over 10} \left({\sigma_l\over L}\right)^{p-1\over 2}.
\end{eqnarray}

The integrated residual error, $E^I_K$, over a period of time is,

\begin{eqnarray}
 E_K^{I}< C_k{U^2\over 10} \int^t_0 \left({\sigma_l\over L}\right)^{{p-1\over 2}}dt.
\end{eqnarray}

Assuming the pair separation scaling $\sigma^2_l\sim t^{\chi_p}$,  where $t$ is the time and for some $\chi_p>0$, yields,

\begin{eqnarray}
 E_K^{I}\lesssim  C_k{UL\over 10}  
                        \left({\sigma_l\over L}\right)^{{{p-1}\over 2}+{{2\over \chi_p}}}.
\end{eqnarray}

If the pair diffusivity scales like, $K\sim \sigma_l^{\gamma_p}$, for some $\gamma_p>0$ then $\chi_p=2/(2-\gamma_p)$ is an exact relation.

The most important quantity is the normalised integrated residual error  with respect to the pair diffusivity,  $e^I_K$. Using the above expression for $\chi_p$, and replacing the scalig with $L$ by scaling with $\eta$,  and absorping all ensuing constants in to a new constant $A_k$, leads to

\begin{eqnarray}
 e_K^{I}= {E^I_K\over K} \lesssim 
{A_k\over \displaystyle 10\left({{\sigma_l\over \eta}}\right)^{2\gamma_p-{{p+3}\over 2}}}.
\end{eqnarray}

For strict locality scaling $\gamma_p=(1+p)/2$, and this becomes

\begin{eqnarray}
 e_K^{I}\lesssim {A_k\over \displaystyle 10\left({\sigma_l\over \eta}\right)^{\gamma_p-1}}.
\end{eqnarray}

Since $\gamma_p-1>0$, $e_K^I$ decreases with increasing pair separation for all $p> 1$. For $p=5/3$, we have $\gamma_p=4/3$ and we obtain,

\begin{eqnarray}
 e_K^{I}\lesssim {A_k\over 10\displaystyle\left({\sigma_l\over \eta}\right)^{1/3}}.
\end{eqnarray}

Thus, as $\sigma_l/\eta\to \infty$ then $e^I_K$ decreases; and as $\sigma_l/\eta\to 0$  then $e^I_K$ increases. 

Even for non-local scaling, assuming that $\gamma_p$ does not deviate too far from the local scaling, the above order of magnitude for $e^I_K$ is still approximately true. In fact, in KS we know, Section 2.3, that $\gamma_p$ is slightly greater than locality so the errors will be slightly smaller than in eqution (25).

{The magnitude of these errors will also depend upon $A_k$; but $A_k$ cannot be determined from theoretical considerations alone, although it is reasonable to assume that $A_k$ is small enough for the $e^I_K$ to be negligible towards larger separations as $\sigma_l/\eta\to \infty$ { -- this assumption will be justified by numerical simulations}. Even so this does not guarantee that the error will remain small at smaller separations, since $e^I_K\to \infty$ as $\sigma_l/\eta\to 0$.

However, {between these two asymptotic limits} there must exist an intermediate range of separations, between $1< \sigma_l/\eta< \infty$, where the errors remain negligible. The crucial question is, how wide is this range of scales? }

If this intermediate range is short then the KS errors will be significant at almost all separations. However, if it is long then the KS errors will be negligble inside this intermediate range where true inertial subrange scaling can be expected to be manifested.

To determine the extent of this intermediate range of scales, if it exists at all, simulations with KS must be performed with very large inertial subranges.

\section{Simulations and Results}

The normalised integrated error, $e^I_K$, is scale dependent and reduces with increasing separation. The KS diffusivity is given by, $K^s\approx K(1 + e_K^I)\to K$ as $\sigma_l/\eta\to \infty$. It is expected that if there is an appreciable intermediate range where the errors are negligible, then the power scaling in $K^s$ must be constant and asymptotic to the limiting case where $\sigma_l/\eta\to \infty$. The extent of this intermediate range is determined by the range over which the power scaling in $K^s$ is constant.

Furthermore, significant levels of the sweeping error means that the fluid particles cut through KS eddies, and therefore must be accompied by high levels of noise -- the larger the relative sweeping error the larger the noise level.

Thus, where the errors are negligible it is expected that the correct power law scaling, $K^s\approx K$, will be observed in that part of the of the inertial subrange.

On the other hand, where the errors are significant it is expected that $K^s$ will deviate from the true power law scaling for $K$ and also be accompanied with significant statistical noise due to fluid particles being swept through local eddies in that part of the inertial subrange. Even in this case, however, it is expected that the errors and the associated noise diminish as the pair separation increases.

\subsection{Frames of reference}
Comparison will be made between two cases: first, where $K^s$ is obtained from KS in the physically correct sweeping frame of reference; and second, the case where large scale random sweeping velocities are explicitly added to the flow.

The analysis for the latter case is similar to that which leads to equation (24), except that the factor of $10$ in the denominator is now $1$. Both of these cases can therefore be written collectively as,

\begin{eqnarray}
 e_K^{I}\lesssim {A_k\over C\displaystyle\left({\sigma_l\over \eta}\right)^{\gamma_p-1}};
\end{eqnarray}

where $C=10$ in the swept frame of reference, and $C=1$ when large random scales are included in the simulations -- the residual errors are an order of magnitude smaller in the swept frame of reference.

{\bf Case 1: Swept frame.} Set the spectrum to be $E(k)\sim k^{-p}$ in the inertial subrange, and set $E(k)=0$ for $k<k_1$.

{\bf Case 2: Non-swept frame.} Set the spectrum to be $E(k)\sim k^{-p}$ in the inertial subrange as in Case 1, and add $E(k)=E_0\delta(k-k_0)$  at some low wavenumber $k_0<k_1$, and such that $E_0$ is the energy in the von Karman spectrum in the range $0<k<k_1$.

A very small fixed timestep, smaller than any timescale in the system $dt\ll \tau_{\eta}$, is used in all the simulations reported here. $\tau_\eta$ is the Kolmogorov time scale.

\begin{figure}
\begin{center}
\mbox{\subfigure{\includegraphics[width=6cm]{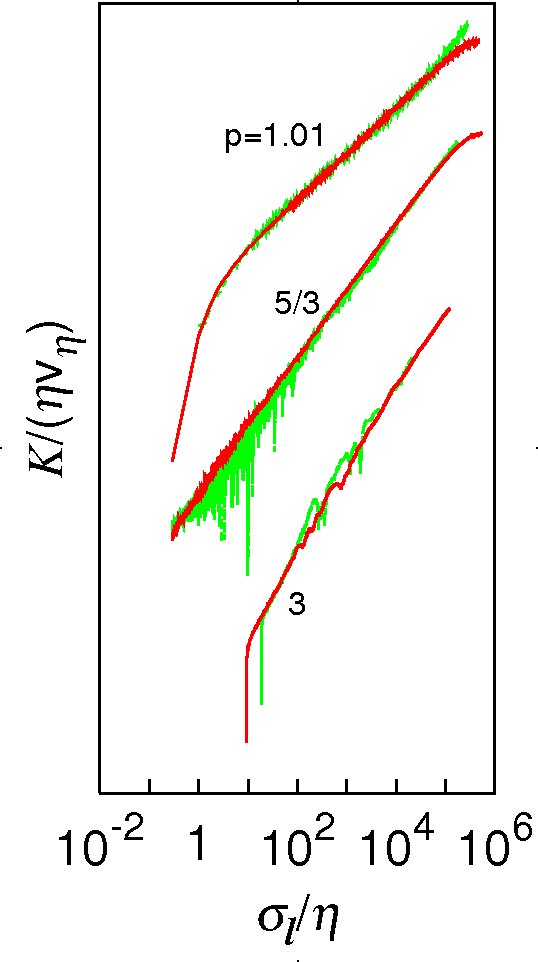}}}
\caption{
\label{fig3} The turbulent diffusivity as $\log(K/(\eta v_\eta))$ against $\log(\sigma_l/\eta)$ obtained from KS. From top to bottom, $p=$ $1.01$, $5/3$, $3$. Case 1 (red lines), Case 2 (green lines).}
\end{center}
\end{figure}

\subsection{Kinematic Simulations}

Kinematic Simulation [4,5] is a Lagrangian method for particle diffusion in which the velocity fileld is prescribed as a sum of energy-weighted Fourier modes. It is akin to the widely used random flight type of statistical models in which the dynamical interactions between turbulent length scales is not explcitly modeled, rather the overall effect on the statistical moments of particle diffusion is mimicked. In KS this is accomplished by specifying the energy spectrum $E(k)$. KS continues to be used in turbulent diffusion studies for both passive and inertial particle motion, including cases with generalized power-law  energy  spectra of the form $E(k)\sim k^{-p}$ for $p>1$, Maxey \cite{Maxey1987}, Turfus \cite{Turfus1987}, Fung \& Vassilicos \cite{Fung1998}, Malik \& Vassilicos \cite{Malik1999}. {Meneguz \& Reeks \cite{Meneguz2011} carried out a DNS of inertial particle motion, and compared it to results from KS which they found to  agree well with the DNS.}

KS generates turbulent-like non-Markovian particle trajectories by releasing particles in flow fields that are incompressible by construction and which satisfy Eulerian statistics up to second order. A turbulent flow field realization is produced as a truncated Fourier series,

\begin{eqnarray}
   {\bf W}({\bf x},t) =  \sum_{n=1}^{N}
   \left[{({\bf A_n\times \hat k_n})\cos{({\bf k_n\cdot x} +\omega_n t)} + 
          ({\bf B_n\times \hat k_n})\sin{({\bf k_n\cdot x} +\omega_n t)} }\right]
\ \ \ \
\end{eqnarray}

where $N$ is a suitable number of representative wavemodes, typically hundreds for very long spectral ranges, $k_\eta/k_1\gg 1$. $\hat {\bf k}_n$ is a random unit vector (${\bf k}_n = \hat {\bf k}_n k_n$ and $k_n = |{\bf k}_n|$). The coefficients ${\bf A}_n$ and ${\bf B}_n$ are chosen such that their orientations are randomly distributed and uncorrelated with any other Fourier coefficient or wavenumber, and their amplitudes are
determined by $\langle {\bf A}^2_n \rangle = \langle {\bf B}^2_n \rangle \propto k_nE(k_n)$, where $E(k),  \ \ k_1\le k\le k_\eta$, is the turbulent energy specturm. The angled brackets $\langle \cdot \rangle$ denotes the ensemble average over many flow fields. This construction ensures incompressibility in each flow realization, ${\bf \nabla} \cdot {\bf u} = 0$. The flow field ensemble generated in this manner is statistically homogeneous, isotropic, and  stationary.

{An important feature of KS is that unlike some other Lagrangian methods, by generating entire kinematic flow fields in which particles are tracked it does not suffer from the crossing-trajectories error which is caused when two fluid particles occupy the same location at the same time in violation of incompressibility; but because KS flow fields are incompressible by construction this error is completely eliminated.}

The energy spectrum $E(k)$ can be chosen  freely within a finite range of scales. In turbulent particle pair studies the interest is in Kolmogorov-like power law spectra,

\begin{eqnarray}
      E(k) &=& C_E \varepsilon^{2/3}L^{5/3-p}k^{-p}, \ \ k_1\le k\le k_\eta (=2\pi/\eta), \ \ 
      1<p\le 3
\end{eqnarray}

$C_E$ is a constant. The largest represented scale of turbulence is $2\pi/k_1$, and the smallest is the Kolmogorov scale $\eta=2\pi/k_\eta$.
The constant is normalized such that the total energy contained in the range 
$k_1\le k \le k_\eta$ is $3(u')^2/2$, where $u'$ is the rms turbulent velocity fluctuations 
in each direction.  $\varepsilon(p)$ is determined by integrating the spectrum, 
$\int_{k_1}^{k_\eta} E(k)dk=3(u')^2/2$. ($p=1$ is a singular limit which is not consider here.)
$v_\eta = (\varepsilon\eta)^{1/3}$ is the velocity micrcoscale, and $\tau_\eta = \varepsilon^{-1/3} \eta^{2/3}$ is the Kolmogorov time micrcoscale. 

The frequencies are chosen according to usual practice to be proportional to the eddy-turnover frequencies, i.e. $\omega_n= \lambda\sqrt{k_n^3E(k_n)}$. The choice of $\lambda$ is somewhat arbitrary, but provided $\lambda<1$ it does not affect the diffusion scaling itself -- even frozen field with $\lambda=0$ yields the same scaling \cite{Malik1996}. $\lambda=0.5$ is a common practice in KS which is also chosen here.

The distrbution of the wavemodes is geometric, $k_n=k_1 r^{n-1}$, with $r=(k_\eta/k_1)^{1/(N-1)}$. The grid size in wavemode-space of  the $n^{th}$ wavemode is $\delta k_n = k_n(\sqrt{r}- 1/\sqrt{r})$.

A particle trajectory is obtained by integrating the Lagrangian velocity ${\bf W}_L(t)$,
\begin{eqnarray}
   {d{\bf x} \over dt} = {\bf W}_L(t) = {\bf W}({\bf x},t).
\end{eqnarray}
Pairs of trajectories are harvested over a large ensemble of flow realizations and pair statistics are then obtained from it for analysis.

{The turbulent difusivity itself can be computed in two ways. Directly from the forumla $K(l) \sim \langle {\bf l}\cdot {\bf v}(l) \rangle$, i.e. the ensemble average of the scalar producted of ${\bf v}$ and ${\bf l}$. But it has been found that using the equivalent formula, $K(l) \sim d\langle l^2\rangle /dt$, i.e. the derivative of the $\langle l^2  \rangle$, converges  faster statistically needing a much smaller ensemble of trajectories, although the two methods give identical results for large enesmbles of particle trajectories. The latter method has been adopted here.}

{Lagrangian statistics are the physically meaningful output from KS. It is {\em not} correct to compare the kinematically generated flow fields directly with DNS flow fields. As such, KS is like Lagrangian methods such as Random Walk models where an individual particle trajectory has no physical meaning, but the ensemble average over many such random trajectories produces physically meaningful Lagrangian statistics.}

\subsection{Results}

KS simulations were performed with $L=1$, $k_1=1/L=1$, and $k_\eta=10^6$, $C_E=1.5$ (Kolmogorov constant) and $u'=1$. There were $200$ wavemodes per realization.

In Case 1 (swept frame of reference), $E(k)=0$ for $k< 1$. 

In Case 2, large scale random sweeping were added at the low wavenumber $k_0=1/10$, with $E(k)=E_0\delta(k-k_0)$. The energy in these sweeping scales, $E_0$, was equal to the energy contained in the von Karman turbulence spectrum for $k<1$.  $k_0$ corresponds approximately to the location of the peak in the von Karman spectrum.

In both cases, three different power spectra were considered with, respectively, $p=1.01, 5/3$, and $3$. 
With 8 pairs released in 5000 flow realizations, the Lagrangian statistics were obtained from 40,000 particle pair trajectories.

Fig. 3 shows log-log plots of the pair diffusivity $K/(\eta v_\eta)$ against $\sigma_l/\eta$, for Case 1 (red lines) and Case 2 (green lines). The energies in the two cases are different, so Case 2 plots have been shifted vertically in order to compare the two cases directly. This does not affect the scalings (the slopes) which is the main interest here. Hence the ordinate is shown without scale.

For $p=1.01$, the two cases align with a constant power-law scaling, $\gamma_{1.01} \approx 1.07$, over most of the inertial subrange of scales and there is very little statistical noise, indicating that $e^I_K(\sigma_l) \ll 1$ at all separations in this part of the inertial subrange in both cases. 
In this limit, locality is very strong, and the relative motion is unaffected by the long range sweeping. 
The obtained slope is indeed very close to the exact locality scaling of $1.005$. 

For $p=5/3$ (Kolmogorov turbulence), in Case 1 (red)  a clear power-law scaling is observed, $\gamma_{Kol} \approx 1.53>4/3$, and very little statistical noise in the range $1<\sigma_l/\eta <10^5$, indicating that $e^I_K(\sigma_l)\ll 1$ in this range of scales. Case 2 (green) deviates increasingly from Case 1 at small inertial separations where it is also accompanied with increasing levels of noise. Nevertheless, the agreement between the two cases for $\sigma_l/\eta>10^2$ is good.

For $p=3$, the two cases overlap with a power-law scaling, $\gamma_{3} = 2$,  with almost no statistical noise. In this limit nearly all the energy is in the large scales and inertial range scaling is no longer applicable; rather uniform strained motion with the characteristic slope of $2$ is obtained.

\section{Discussion and Conclusions}

All the results in Fig. 3 are consistent with the numerical analysis and the theoretical predictions in section 2. The constant power law scaling over most of the inertial subrange of separations, $1<\sigma_l/\eta<10^5$, and the very low level of statistical noise  in the swept frame of reference are especially important. (The departure from this for $\sigma_l/\eta<1$ observed in Fig. 3 is outside of the inertial subrange.)

The KS sweeping errrors in this frame of reference is thererfore negligible for most practical purposes. It is possible that KS could produce negligible sweeping errors in even bigger intermediate ranges than reported here, but the current simulations are the maximum size of inertial subrange possible, $k_\eta/k_1=10^6$, with double-precision accuracy.

It is remarkable that even when large scale sweeping is included, the KS sweeping errors remain small in the range $\sigma_l/\eta>10^2$.

{It is also noted that some Direct Numerical Simulation (DNS) show pair diffusion which appear to display locality scaling, see \cite{Biferale2014} for example. However, the maximum inertial range obtained in DNS to date is around $k_\eta/k_1\approx 10^2$, which is much shorter than required to test pair diffusion scaling reliably -- that requires $k_\eta/k_1> 10^4$. Thus the current KS results cannot be compared directly with DNS at the present time. However, for low Reynolds KS has been validated against DNS for turbulent pair diffusion by Malik \& Vassilicos \cite{Malik1999}; here not only did the pair diffusion from KS closely match the DNS results with the same energy specturm, but  the fourth order statistic, the kurtosis in the pair separation, also matched remarkably well.}

{The main contribution of this work is that it has been demonstrated that in a reference frame moving with the large energy containing scales the sweeping errors in the turbulent pair diffusion process in KS is negligible in the inertial subrange where, $1<\sigma_l/\eta<10^5$. This is significant not only because it {amends the  previous theory of \cite{TD2005,Nicolleau2011,Eyink2013} , but it is important for turbulent diffusion modeling and applications in general, and especially for pair diffusion studies.}}

If the sweeping errors are negligible, why is locality scaling not oberved for $p=5/3$ where KS yields $\gamma_{Kol}\approx 1.53 >4/3$? There are two possible explanations. First, other errors could be present in KS, not due to the sweeping, but to as yet unknown sources; but no one has proposed any such source of error in, and so this remains very speculative. 

The second possibility is that the locality hypothesis itself may be error. This goes against the widely accepted theory of Richardson [1], and no one has proposed alternative theory. On the other hand, it should be noted that the locality scaling for the pair diffusion has never been confirmed unequivocally as noted by Salazar \& Collins \cite{Salazar2009} -- so there is room for new thinking in this field.

These are currently the subjects of active research by the author. What can be said for certain at the present time is that the departure from locality scaling observed in KS cannot be attributed to the sweeping effect.

\section*{Acknowledgments}

The author would like to thank SABIC for funding this work through project number SB101011, and the ITC Department at KFUPM for making available the High Performance Computing facility for this project. 

\nolinenumbers

%
%
%


\begin{thebibliography}{14}

{
\bibitem[1]{Richardson1926}
{\sc Richardson L. F.}  Atmospheric diffusion shown on a distance-neighbour graph.
{\em Proc. Roy. Soc. Lond. A\/} {100}, 709-737 (1926).

\bibitem[2]{Obukhov1941}
{\sc Obukhov, A.} Spectral energy distribution in a turbulent flow.
{\em Izv. Akad. Xauk. SSSR. Ser. Geogr. i Geojz\/} {5}, 453--466. (Translation : Ministry of Supply. p. 21 1097) (1941).

\bibitem[3]{Morel1974}
{\sc Morel, A. \& Larchaveque, M.} Relative Dispersion of Constant-Level Balloons in the 
200-mb General Circulation.
{\em J. Atm. Sci.\/} {31}, 2189-2196 (1974).

\bibitem[4]{Kraichnan1970}
{\sc Kraichnan, R. H.} Diffusion by a random velocity field.
{\em Phys. Fluids\/} {13}, 22--31 (1970).

\bibitem[5]{Fung1992}
{\sc Fung J. C. H., Hunt J. C. R.,  Malik N. A. \& Perkins R. J.} 
Kinematic simulation of homogeneous turbulence by unsteady random Fourier modes.
{\em J.~Fluid Mech.\/} {236}, 281--318 (1992).

\bibitem[6]{TD2005} 
{\sc Thomson, D. J. \& Devenish, B. J.} 
Particle pair separation in kinematic simulations.
{\em J.~Fluid Mech.\/} {526}, 277--302 (2005).

\bibitem[7]{Nicolleau2011}
{\sc Nicolleau, F. C. G. A. \& Nowakowski, A. F.} 
Presence of a Richardson's regime in kinematic simulations.
{\em Phys. Rev. E\/} {83}, 056317 (2011).

\bibitem[8]{Eyink2013}
{\sc Eyink, G. L. \& Benveniste, D.} 
Suppression of particle dispersion by sweeping effects in synthetic turbulence.
{\em Phys. Rev. E\/} {87}, 023011 (2013).

\bibitem[9]{Batchelor1952}
{\sc Batchelor, G. K.} Diffusion in field of Homogeneous Turbulence II. The relative motion of particles.
{\em Math. Proc. Camb. Phil. Soc.\/} {48(2)}, 345-362, (1952).

\bibitem[10]{Maxey1987}
{\sc Maxey, M. R.} The gravitational settling of aerosol particles in homogeneous turbulence and random flow fields.
{\em J. Fluid Mech.\/} {174}, 441-465 (1987).

\bibitem[11]{Turfus1987}
{\sc Turfus, C. \& Hunt, J. C. R. }
A stochastic analysis of the displacements of fluid element in inhomogeneous turbulence using Kraichnan's method of random modes.
{\em Advances in Turbulence (ed. G. Comte-Bellot \& J. Mathieu)\/} {Springer}, 191-203 (1987).

\bibitem[12]{Fung1998}
{\sc Fung, J. C. H.  and Vassilicos, J.C. } Two-particle dispersion in turbulent-like flows.
{\em Phys. Rev. E\/} {57}, 1677 (1998).

\bibitem[13]{Malik1999}
{\sc Malik, N. A. \& Vassilicos, J. C.}
A Lagrangian model for turbulent dispersion with turbulent-like flow structure: comparison with direct numerical simulation for two-particle statistics.
{\em Phys. Fluids\/} {11}, 1572-1580 (1999).

\bibitem[14]{Meneguz2011}
{\sc [Meneguz, E. \& Reeks, M. W.}
Statistical properties of particle segregation in homogeneous isotropic turbulence.
{\em J. Fluid Mech.\/} {686}, 338-351 (2011).

{
\bibitem[15]{Malik1996}
{\sc Malik N. A.} Structural diffusion in 2D and 3D random flows.
{\em Adv. in Turbulence\/} {\bf VI}, eds. S. Gavrilakis et al., 619--620 (1996).

\bibitem[16]{Biferale2014}
{\sc Biferale L., Lanotte A. S., Scatamacchia R., \& Toschi F.} 
Extreme events for two-particles separations in turbulent flows. 
{\em Prog. in Turbulence\/} {\bf V (149)}, 
Springer Proceedings in Physics, 9–16. Springer, 2014.
}

\bibitem[17]{Salazar2009}
{\sc Salazar J. P. L. \& Collins L. R.} Two-Particle Dispersion in Isotropic Turbulent Flows.
{\em Annu. Rev.  Fluid Mech.\/} {\bf 41}, 405--432 (2009).  
}


\end{thebibliography}
\end{document}